\documentclass[prd,reprint,showpacs,showkeys]{revtex4-1}
\usepackage{amsfonts}
\usepackage{amssymb}
\usepackage{amsmath}
\usepackage{graphicx}
\usepackage[font={footnotesize,it}]{caption}

\begin{document}

\title{Stability of generic cylindrical thin shell wormholes}
\author{S. Habib Mazharimousavi}
\email{habib.mazhari@emu.edu.tr}
\author{M. Halilsoy}
\email{mustafa.halilsoy@emu.edu.tr}
\author{Z. Amirabi}
\email{zahra.amirabi@emu.edu.tr}
\affiliation{Department of Physics, Eastern Mediterranean University, G. Magusa, north
Cyprus, Mersin 10, Turkey. }
\date{\today }

\begin{abstract}
We revisit the stability analysis of cylindrical thin shell wormholes which
have been studied in literature so far. Our approach is more systematic and
in parallel to the method which is used in spherically symmetric thin shell
wormholes. The stability condition is summarized as the positivity of the
second derivative of an effective potential at the equilibrium radius, i.e. $%
V^{\prime \prime }\left( a_{0}\right) >0$. This may serve as the master
equation in all stability problems for the cylindrical thin-shell wormholes.
\qquad
\end{abstract}

\pacs{04.70.Bw; 04.40.Nr; 04.20.Gz}
\keywords{Thin-shell wormhole; Stability; Cylindrically symmetric; }
\maketitle

\section{Introduction}

Upon breaking of spherical symmetry in an axial direction we arrive at
cylindrical symmetry. A large number of systems fail to satisfy spherical
symmetry and are considered within the context of cylindrical (or axial)
symmetry. Spacetimes that depend on radial $r$ and time $t$ are known to
describe cylindrical waves. Replacing $t$ with the spacelike coordinate $z$
gives rise to static axially symmetric spacetimes. Our interest in this
study is to suppress the $t$ and $z$ dependences and consider spacetimes
depending only on the radial $r$ coordinate. This amounts to admit three
Killing vectors $\xi _{t}^{\mu },$ $\xi _{z}^{\mu }$ and $\xi _{\varphi
}^{\mu }$ in the Weyl coordinates $\{t,r,z,\varphi \}$. Historically the
first such example was given by Levi-Civita \cite{LC}. Topological defect
spacetimes believed to form during the early universe such as cosmic strings 
\cite{Vilenkin} also fit into this class. The latter's current-like source
is located along an axis which creates a deficit angle in the surrounding
space so that it gives rise to gravitational lensing. Still another example
for cylindrically symmetric metrics which is powered by a beam-like magnetic
field is the Melvin's magnetic universe \cite{Melvin}. Addition of extra
fields such as Brans-Dicke scalar or various electromagnetic fields to the
cylindrical metrics has been extensively searched in the literature \cite%
{BransDicke}. Recently we have given an example of Weyl solution in which
the magnetic Melvin and Bertotti-Robinson metrics are combined in a simple
Einstein-Maxwell metric \cite{MH1}. There is already a large literature
related to the spherically symmetric thin-shell wormholes (TSW) \cite{TSWS}
but for the cylindrically symmetric cases the published literature is
relatively less \cite{TSWC}. From this token we wish to consider a general
class of cylindrically symmetric spacetimes in which the metric functions
depend only on the radial function $r$ to construct TSWs. As usual, our
method is cutting and pasting of two cylindrically symmetric spacetimes
which unlike the spherical symmetric cases are more restrictive toward
asymptotic flatness. Being $z-$independent the metric is same for both $z=0$
and $\left\vert z\right\vert =\infty $. Yet the areal/ radial flare-out
conditions must be satisfied \cite{BR, ES1, ES2, R}, in spite of the fact
that the spacetime may not be asymptotically flat. The radial ($P_{r}$) and
axial ($P_{z}$) pressures are assumed to be functions of the energy (mass)
density $\sigma $. The junction conditions at the intersection determine the
throat equation as a function of the proper time. From the extrinsic
curvature components we extract an energy equation for a one-dimensional
particle of the form $\ \dot{a}^{2}+V\left( a\right) =0,$ where $a\left(
\tau \right) $ is the radius of the throat and dot means a proper time
derivative. The form of the potential $V\left( a\right) $ can be rather
complicated but since we are interested in the stability we need to
investigate only the second derivative of the potential around the
equilibrium radius of the throat. The parametric plotting of the second
derivative of the potential $V^{\prime \prime }\left( a_{0}\right) >0$,
where $a_{0}$ is the equilibrium radius, reveals the stability region for
the TSW under consideration. Our perturbation addresses only to the radial
and linear cases for which we may adopt Equations of State (EoS) for the
surface energy-momentum at the throat. Adding extra source amounts to the
fact that the covariant divergence of the surface energy-momentum is
non-zero. The structural equations for perturbations expectedly are more
complicated than the spherical symmetric case, which is natural from the
less symmetry arguments. Concerning the exotic/ normal matter, however, our
formalism does not add anything new, i.e. our matter to thread the TSW is
still exotic. In a recent study we have proposed that in order to get
anything total but exotic matter as source, albeit locally is exotic the
geometry of the throat must be of prolate/ oblate type \cite{MH2}.

Organization of the paper is as follows. In Section II we consider a general
line element with cylindrical symmetry and derive the stability condition
for the TSW. In Sec. III we make applications of the result found in Sec.
II. We complete the paper with Conclusion in Sec. IV.

\section{General analysis for cylindrically symmetric TSW}

Let's consider two static, cylindrically symmetric spacetimes $\mathcal{M}%
_{\pm }$ \cite{BR, GLV}, in Weyl coordinates%
\begin{multline}
ds^{2}=-e^{2\gamma _{\pm }\left( r_{\pm }\right) }dt_{\pm }^{2}+e^{2\alpha
_{\pm }\left( r_{\pm }\right) }dr_{\pm }^{2}+ \\
e^{2\xi _{\pm }\left( r_{\pm }\right) }dz_{\pm }^{2}+e^{2\beta _{\pm }\left(
r_{\pm }\right) }d\varphi _{\pm }^{2}.
\end{multline}%
By gluing these two manifolds at their boundaries $\Sigma _{\pm }$, one can,
in principle, make a single complete manifold. Each separate spacetime $%
\mathcal{M}_{\pm }$ must satisfy the Einstein's equations with a general
form of energy momentum tensor, $T_{\mu \pm }^{\nu }=\left[ -\rho _{\pm
},p_{r\pm },p_{z\pm },p_{\varphi \pm }\right] $%
\begin{equation}
G_{\mu \pm }^{\nu }=T_{\mu \pm }^{\nu }
\end{equation}%
with the unit convention ($8\pi G=c=1$). Einstein's equations in each
spacetime admit (for simplicity we suppress sub $\pm $ for each spacetime
but they are implicitly there)%
\begin{equation}
-\rho =e^{-2\alpha }\left[ \beta ^{\prime \prime }+\xi ^{\prime \prime
}+\beta ^{\prime 2}+\left( \xi ^{\prime }-\alpha ^{\prime }\right) \left(
\beta ^{\prime }+\xi ^{\prime }\right) \right]
\end{equation}%
\begin{equation}
p_{r}=e^{-2\alpha }\left[ \left( \beta ^{\prime }+\xi ^{\prime }\right)
\gamma ^{\prime }+\xi ^{\prime }\beta ^{\prime }\right]
\end{equation}%
\begin{equation}
p_{z}=^{-2\alpha }\left[ \gamma ^{\prime \prime }+\beta ^{\prime \prime
}+\gamma ^{\prime 2}+\left( \gamma ^{\prime }+\beta ^{\prime }\right) \left(
\beta ^{\prime }-\alpha ^{\prime }\right) \right] ,
\end{equation}%
and%
\begin{equation}
p_{\varphi }=^{-2\alpha }\left[ \gamma ^{\prime \prime }+\xi ^{\prime \prime
}+\gamma ^{\prime 2}+\left( \gamma ^{\prime }+\xi ^{\prime }\right) \left(
\xi ^{\prime }-\alpha ^{\prime }\right) \right] ,
\end{equation}%
in which a prime stands for the derivative with respect to $r_{\pm }$
depending on the manifold under consideration.

After gluing the two spacetimes at their boundaries whose equation, in our
study, is given by $\mathcal{H}=r-a\left( \tau \right) =0$ the intrinsic
line element on the common boundary $\Sigma =\Sigma _{\pm }$ can be written
as%
\begin{equation}
ds_{\Sigma }=-d\tau ^{2}+e^{2\xi \left( a\right) }dz^{2}+e^{2\beta \left(
a\right) }d\varphi ^{2}
\end{equation}%
in which $a=a\left( \tau \right) $ is a function of proper time $\tau $. The
normal $4-$vector on the timelike hypersurface $\Sigma $ is defined as%
\begin{equation}
n_{\gamma }^{\left( \pm \right) }=\left( \pm \left\vert g^{\alpha \beta }%
\frac{\partial \mathcal{H}}{\partial x^{\alpha }}\frac{\partial \mathcal{H}}{%
\partial x^{\beta }}\right\vert ^{-1/2}\frac{\partial \mathcal{H}}{\partial
x^{\gamma }}\right) _{\Sigma },
\end{equation}%
which in closed form becomes%
\begin{equation}
n_{\gamma }^{\left( \pm \right) }=\pm \left( -e^{\alpha _{\pm }+\gamma _{\pm
}}\dot{a},e^{2\alpha _{\pm }}\sqrt{\Delta _{\pm }},0,0\right) _{\Sigma }
\end{equation}%
where $\Delta _{\pm }=e^{-2\alpha _{\pm }}+\dot{a}^{2}$ and a dot stands for
the derivative with respect to the proper time. Next, we find the extrinsic
curvature on the hypersurface $\Sigma $ defined as%
\begin{equation}
K_{ij}^{\left( \pm \right) }=-n_{\gamma }^{\left( \pm \right) }\left( \frac{%
\partial ^{2}x_{\pm }^{\gamma }}{\partial X_{\pm }^{i}\partial X_{\pm }^{j}}%
+\Gamma _{\pm \alpha \beta }^{\gamma }\frac{\partial x_{\pm }^{\alpha }}{%
\partial X_{\pm }^{i}}\frac{\partial x_{\pm }^{\beta }}{\partial X_{\pm }^{j}%
}\right) _{\Sigma }
\end{equation}%
in which $X_{\pm }^{i}\in \left\{ \tau ,z_{\pm },\varphi _{\pm }\right\} $
while $x_{\pm }^{\gamma }=\left\{ t_{\pm },r_{\pm },z_{\pm },\varphi _{\pm
}\right\} .$ Explicit calculations yield%
\begin{equation}
K_{\tau }^{\tau \left( \pm \right) }=\pm \left( \frac{1}{\sqrt{\Delta _{\pm }%
}}\left( \ddot{a}+\left( \alpha _{\pm }^{\prime }+\gamma _{\pm }^{\prime
}\right) \dot{a}^{2}+e^{-2\alpha _{\pm }}\gamma _{\pm }^{\prime }\right)
\right) _{\Sigma },
\end{equation}%
\begin{equation}
K_{z}^{z\left( \pm \right) }=\pm \left( \xi _{\pm }^{\prime }\sqrt{\Delta
_{\pm }}\right) _{\Sigma }
\end{equation}%
and%
\begin{equation}
K_{\varphi }^{\varphi \left( \pm \right) }=\pm \left( \beta _{\pm }^{\prime }%
\sqrt{\Delta _{\pm }}\right) _{\Sigma }.
\end{equation}%
By considering a standard energy momentum on the shell i.e., $%
S_{i}^{j}=diag.\left( -\sigma ,P_{z},P_{\varphi }\right) ,$ the Israel
junction conditions \cite{Israel} imply%
\begin{equation}
\left\langle K_{i}^{j}\right\rangle -\left\langle K\right\rangle \delta
_{i}^{j}=-S_{i}^{j}
\end{equation}%
in which $\left\langle K_{i}^{j}\right\rangle =\left\langle
K_{i}^{j}\right\rangle _{+}-\left\langle K_{i}^{j}\right\rangle _{-}$ and $%
\left\langle K\right\rangle =\left\langle K_{i}^{i}\right\rangle .$ The
latter amounts to%
\begin{equation}
\sigma =-\left[ \left( \xi _{+}^{\prime }+\beta _{+}^{\prime }\right) \sqrt{%
\Delta _{+}}+\left( \xi _{-}^{\prime }+\beta _{-}^{\prime }\right) \sqrt{%
\Delta _{-}}\right] ,
\end{equation}%
\begin{multline}
P_{z}=\frac{\left( \ddot{a}+\left( \alpha _{+}^{\prime }+\gamma _{+}^{\prime
}\right) \dot{a}^{2}+e^{-2\alpha _{+}}\gamma _{+}^{\prime }\right) }{\sqrt{%
\Delta _{+}}}+ \\
\frac{\left( \ddot{a}+\left( \alpha _{-}^{\prime }+\gamma _{-}^{\prime
}\right) \dot{a}^{2}+e^{-2\alpha _{-}}\gamma _{-}^{\prime }\right) }{\sqrt{%
\Delta _{-}}}+\beta _{+}^{\prime }\sqrt{\Delta _{+}}+\beta _{-}^{\prime }%
\sqrt{\Delta _{-}}
\end{multline}%
and%
\begin{multline}
P_{\varphi }=\frac{\left( \ddot{a}+\left( \alpha _{+}^{\prime }+\gamma
_{+}^{\prime }\right) \dot{a}^{2}+e^{-2\alpha _{+}}\gamma _{+}^{\prime
}\right) }{\sqrt{\Delta _{+}}}+ \\
\frac{\left( \ddot{a}+\left( \alpha _{-}^{\prime }+\gamma _{-}^{\prime
}\right) \dot{a}^{2}+e^{-2\alpha _{-}}\gamma _{-}^{\prime }\right) }{\sqrt{%
\Delta _{-}}}+\xi _{+}^{\prime }\sqrt{\Delta _{+}}+\xi _{-}^{\prime }\sqrt{%
\Delta _{-}}.
\end{multline}%
To complete this section we consider a thin-shell on the junction which is
constructed by\ the same bulk spacetime so that on the boundaries $\gamma
,\alpha ,\beta $ and $\xi $ are continuous and consequently 
\begin{equation}
\sigma =-2\left( \xi ^{\prime }+\beta ^{\prime }\right) \sqrt{\Delta },
\end{equation}%
\begin{equation}
P_{z}=\frac{2\left( \ddot{a}+\left( \alpha ^{\prime }+\gamma ^{\prime
}\right) \dot{a}^{2}+e^{-2\alpha }\gamma ^{\prime }\right) }{\sqrt{\Delta }}%
+2\beta ^{\prime }\sqrt{\Delta }
\end{equation}%
and%
\begin{equation}
P_{\varphi }=\frac{2\left( \ddot{a}+\left( \alpha ^{\prime }+\gamma ^{\prime
}\right) \dot{a}^{2}+e^{-2\alpha }\gamma ^{\prime }\right) }{\sqrt{\Delta }}%
+2\xi ^{\prime }\sqrt{\Delta }.
\end{equation}%
From now on our study will be concentrated on the case of TSW made from a
single bulk metric.

\subsection{Energy conservation identity}

The energy conservation identity can be found by calculating $S_{;j}^{ij}.$
Our explicit calculations show that 
\begin{equation}
S_{;j}^{ij}\overset{i=\tau }{=}\frac{d\sigma }{d\tau }+\dot{a}\left( \xi
^{\prime }+\beta ^{\prime }\right) \sigma +\dot{a}\left( \xi ^{\prime
}P_{z}+\beta ^{\prime }P_{\varphi }\right) .
\end{equation}%
Furthermore, when we consider the exact form of $\sigma ,P_{z}$ and $%
P_{\varphi }$ we find from the latter%
\begin{equation}
\frac{d\left( \mathcal{A}\sigma \right) }{d\tau }+e^{\beta }P_{z}\frac{%
d\left( e^{\xi }\right) }{d\tau }+e^{\xi }P_{\varphi }\frac{d\left( e^{\beta
}\right) }{d\tau }=\dot{a}\mathcal{A}\Xi
\end{equation}%
in which 
\begin{equation}
\Xi =\sigma \left[ \frac{\beta ^{\prime 2}+\xi ^{\prime 2}+\beta ^{\prime
\prime }+\xi ^{\prime \prime }}{\beta ^{\prime }+\xi ^{\prime }}-\left(
\alpha ^{\prime }+\gamma ^{\prime }\right) \right] ,
\end{equation}%
and the surface area of the shell $\mathcal{A}=e^{\beta +\xi }.$ In Eq. (22) 
$\frac{d\left( \mathcal{A}\sigma \right) }{d\tau }$ is the time change of
the total internal energy of the shell, $e^{\beta }P_{z}\frac{d\left( e^{\xi
}\right) }{d\tau }$, $e^{\xi }P_{\varphi }\frac{d\left( e^{\beta }\right) }{%
d\tau }$ and $\dot{a}\mathcal{A}\Xi $ are the works done in $z,$ $\varphi $
directions and external forces, respectively. This is comparable with the
similar result in spherically symmetric TSW given in \cite{GLV}.

\subsection{Stability of the thin-shell wormhole}

In this section we apply a linear perturbation and investigate whether the
wormhole is stable against the perturbation analysis or not. Our main
assumption is that the matter which supports the TSW obeys the energy
conservation identity. This in turn implies that from (22), 
\begin{equation}
\left( \mathcal{A}\sigma \right) ^{\prime }+e^{\beta }P_{z}\left( e^{\xi
}\right) ^{\prime }+e^{\xi }P_{\varphi }\left( e^{\beta }\right) ^{\prime }=%
\mathcal{A}\Xi
\end{equation}%
in which a prime stands for derivative with respect to $a.$ Following our
linear perturbation the wormhole is dynamic and from (18) one finds the
equation of the throat as a one dimensional motion $\dot{a}^{2}+V\left(
a\right) =0$ with potential 
\begin{equation}
V\left( a\right) =e^{-2\alpha }-\frac{1}{4}\left( \frac{\sigma }{\xi
^{\prime }+\beta ^{\prime }}\right) ^{2}.
\end{equation}%
If $a=a_{0}$ is considered as an equilibrium point with $\dot{a}_{0}=0=\ddot{%
a}_{0}$ then $V\left( a\right) $ can be expanded about the equilibrium point
at which $V\left( a_{0}\right) =0=V^{\prime }\left( a_{0}\right) .$ Also the
components of the energy momentum tensor on the shell when the equilibrium
state is considered are given by%
\begin{equation}
\sigma _{0}=-2\left( \xi _{0}^{\prime }+\beta _{0}^{\prime }\right)
e^{-\alpha _{0}},
\end{equation}%
\begin{equation}
P_{z0}=2\left( \gamma _{0}^{\prime }+\beta _{0}^{\prime }\right) e^{-\alpha
_{0}}
\end{equation}%
and%
\begin{equation}
P_{\varphi 0}=2\left( \gamma _{0}^{\prime }+\xi _{0}^{\prime }\right)
e^{-\alpha _{0}}.
\end{equation}%
We note that a sub-$0$ notation implies that the corresponding quantity is
calculated at the equilibrium point. Next, we find $V^{\prime \prime }\left(
a_{0}\right) =V_{0}^{\prime \prime }$ to examine the motion of the throat.
If $V_{0}^{\prime \prime }>0$ then the motion is oscillatory and the
equilibrium at $a_{0}$ is stable, otherwise it is unstable. To find $%
V^{\prime \prime }$ we need $\sigma ^{\prime }$ and $\sigma ^{\prime \prime
} $ which are given by the energy conservation identity, i.e., 
\begin{equation}
\sigma ^{\prime }=\Xi -P_{z}\xi ^{\prime }-P_{\varphi }\beta ^{\prime
}-\left( \beta ^{\prime }+\xi ^{\prime }\right) \sigma
\end{equation}%
and 
\begin{multline}
\sigma ^{\prime \prime }=\Xi ^{\prime }-P_{z}^{\prime }\xi ^{\prime
}-P_{z}\xi ^{\prime \prime }-P_{\varphi }^{\prime }\beta ^{\prime }- \\
P_{\varphi }\beta ^{\prime \prime }-\left( \beta ^{\prime \prime }+\xi
^{\prime \prime }\right) \sigma - \\
\left( \beta ^{\prime }+\xi ^{\prime }\right) \left( \Xi -P_{z}\xi ^{\prime
}-P_{\varphi }\beta ^{\prime }-\left( \beta ^{\prime }+\xi ^{\prime }\right)
\sigma \right) .
\end{multline}%
Our extensive calculation eventually yields 
\begin{widetext}
\begin{multline}
V_{0}^{\prime \prime }=-\frac{2e^{-2\alpha _{0}}}{\beta _{0}^{\prime }+\xi
_{0}^{\prime }}\left[ \beta _{0}^{\prime 2}\phi _{0}^{\prime }\alpha
_{0}^{\prime }+\left( \alpha _{0}^{\prime }\left[ \phi _{0}^{\prime }+\psi
_{0}^{\prime }+2\right] \xi _{0}^{\prime }+\alpha _{0}^{\prime }\gamma
_{0}^{\prime }-\left[ \beta _{0}^{\prime \prime }+\xi _{0}^{\prime \prime }%
\right] \phi _{0}^{\prime }-\xi _{0}^{\prime \prime }-\gamma _{0}^{\prime
\prime }\right) \beta _{0}^{\prime }\right.  \\
+\left. \left( \xi _{0}^{\prime }\psi _{0}^{\prime }\alpha _{0}^{\prime
}+\alpha _{0}^{\prime }\gamma _{0}^{\prime }-\left( \beta _{0}^{\prime
\prime }+\xi _{0}^{\prime \prime }\right) \psi _{0}^{\prime }-\gamma
_{0}^{\prime \prime }-\beta _{0}^{\prime \prime }\right) \xi _{0}^{\prime }
\right] .
\end{multline}
\end{widetext}Here in this expression, the EoS is considered to be $%
P_{z}=\psi \left( \sigma \right) $, $P_{\varphi }=\phi \left( \sigma \right)
.$ We also note that a prime on a function denotes derivative with respect
to its argument for instance $\psi _{0}^{\prime }=\left. \frac{\partial \psi 
}{\partial \sigma }\right\vert _{\sigma =\sigma _{0}}$ while $\beta
_{0}^{\prime }=\left. \frac{\partial \beta }{\partial a}\right\vert
_{a=a_{0}}.$ Having the form of the metric functions and the EoS are enough
to check whether the TSW is stable or not.

Before we proceed to examine the stability of the TSW, we would like to
introduce the conditions which should be satisfied for having wormhole in
cylindrical symmetry. These conditions were studied in \cite{BR} where the
first condition is called \textit{areal flare-out condition }stating that $%
e^{\xi +\beta }$ must be an increasing function at the throat \cite{ES1,
ES2, R, BR}. The second condition implies $e^{\beta }$ must be an increasing
function at the throat and is called \textit{radial flare-out condition} 
\cite{ES1, ES2, R, BR}. According to \cite{BR} the appropriate condition
would be the \textit{radial flare-out condition. }

\subsection{The Levi-Civita Metric}

Before we find some applications for the general equation (31) among the
known cylindrical TSWs in the literature, in this part, we give the simplest
cylindrical TSW which can be made in the vacuum Levi-Civita (LC) Metric \cite%
{LC}. The LC metric with two essential parameters $b$ and $\delta $ can be
written as%
\begin{equation}
ds^{2}=-br^{4\delta }dt^{2}+r^{4\delta \left( 2\delta -1\right) }\left(
dr^{2}+dz^{2}\right) +\frac{r^{2\left( 1-2\delta \right) }}{b}d\varphi ^{2},
\end{equation}%
in which $b$ is related to the topology of the spacetime giving rise to a
deficit angle $\theta =2\pi \left( 1-\frac{1}{\sqrt{b}}\right) $ \cite{JSD}.
For a physical interpretation of $\delta $ we refer to the third and fourth
papers in Ref. \cite{LC}. Comparing LC line element with our general line
element (1), we find that 
\begin{equation}
e^{2\gamma }=br^{4\delta },\text{ \ }e^{2\alpha }=e^{2\xi }=r^{4\delta
\left( 2\delta -1\right) },\text{ \ }e^{2\beta }=\frac{r^{2\left( 1-2\delta
\right) }}{b}.
\end{equation}%
Once more we note that, in the copy-paste method we consider two copies of
the bulk spacetime (here LC) in which from each we cut the region $r<a\left(
\tau \right) $ and then we join them at $r=a\left( \tau \right) $ to have a
complete manifold. Therefore the outer region of the wormhole is still LC
spacetime with the mentioned essential parameters. Furthermore, the \textit{%
radial} \textit{flare-out condition is satisfied }only for $\delta \leq 
\frac{1}{2}$ while the \textit{areal flare-out condition} is satisfied for
all $\delta .$ Considering the TSW at $r=a\left( \tau \right) $ and using
the general condition of stability i.e., $V_{0}^{\prime \prime }>0$ together
with a linear EoS $\psi ^{\prime }\left( \sigma \right) =\phi ^{\prime
}\left( \sigma \right) =\eta _{0}$ in which $\eta _{0}$ is a constant, we
find%
\begin{equation}
\left( -2\eta _{0}\delta ^{2}+\left( 2\eta _{0}+1\right) \delta -\frac{\eta
_{0}}{2}\right) \left( \delta ^{2}-\frac{1}{2}\delta +\frac{1}{4}\right)
\geq 0.
\end{equation}%
In Fig. 1 we plot the stability region in terms of the parameters $\delta $
and $\eta _{0}$ and as it is observed from the figure, the stability is
sensitive with respect to $\delta .$ In particular for $\delta <\frac{1}{4}$
the stability is not strong enough while for $\frac{1}{4}<\delta <1$ it is
quite strong. Note that the topological parameter $b$ does not play role in
the stability of the LC wormhole.

\begin{figure}[h]
\includegraphics[width=70mm,scale=0.7]{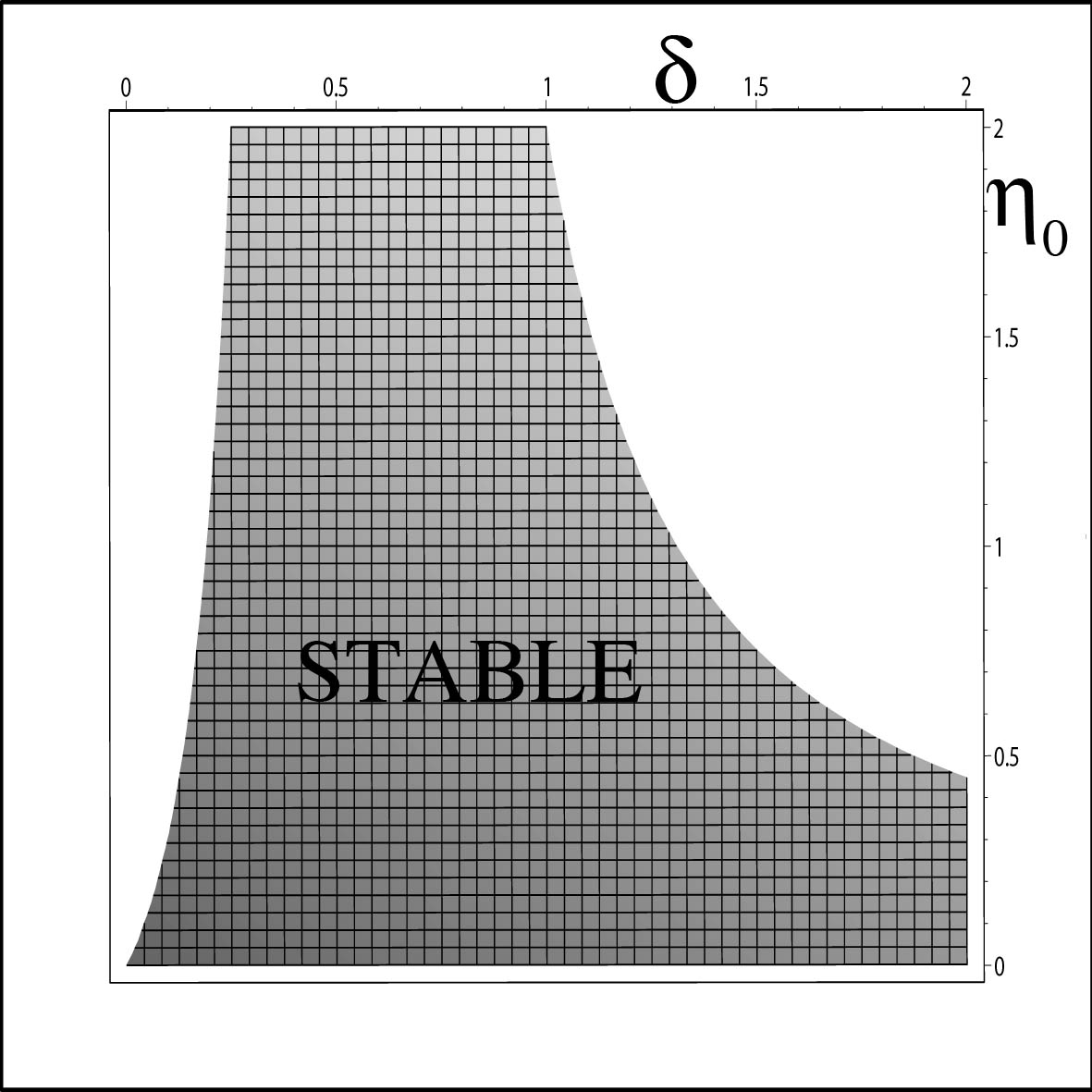}
\caption{Stability of TSW in LC spacetime, supported by a linear gas in
terms of $\protect\delta $ and $\protect\eta _{0}$. We see that the
stability depends on one of the essential parameter i.e. $\protect\delta $
in LC metric. Also the radial flare-out condition is satisfied only for $%
\protect\delta \leq \frac{1}{2}.$}
\end{figure}

\section{Applications}

Eiroa and Simeone in \cite{ES1} have considered a general static cylindrical
metric in $3+1-$dimensions given by%
\begin{equation}
ds^{2}=B\left( r\right) \left( -dt^{2}+dr^{2}\right) +C\left( r\right)
d\varphi ^{2}+D\left( r\right) dz^{2}
\end{equation}%
in which $B\left( r\right) ,$ $C\left( r\right) $ and $D\left( r\right) $
are only function of $r.$ Using the results found above together with $%
e^{2\alpha }=e^{2\gamma }=B,$ $e^{2\beta }=C$ and $e^{2\xi }=D$ one finds

\begin{equation}
\sigma =-\left( \frac{D^{\prime }}{D}+\frac{C^{\prime }}{C}\right) \sqrt{%
\Delta }
\end{equation}%
\begin{equation}
P_{z}=\frac{1}{\sqrt{\Delta }}\left[ 2\ddot{a}+\frac{2B^{\prime }}{B}\dot{a}%
^{2}+\frac{B^{\prime }}{B^{2}}+\frac{C^{\prime }}{C}\Delta \right] ,
\end{equation}%
and%
\begin{equation}
P_{\varphi }=\frac{1}{\sqrt{\Delta }}\left[ 2\ddot{a}+\frac{2B^{\prime }}{B}%
\dot{a}^{2}+\frac{B^{\prime }}{B^{2}}+\frac{D^{\prime }}{D}\Delta \right] .
\end{equation}%
At the equilibrium surface i.e., $a=a_{0}$ we have 
\begin{equation}
\sigma _{0}=-\left( \frac{D_{0}^{\prime }}{D_{0}}+\frac{C_{0}^{\prime }}{%
C_{0}}\right) \frac{1}{\sqrt{B_{0}}}
\end{equation}%
\begin{equation}
P_{z0}=\sqrt{B_{0}}\left[ 2\ddot{a}+\frac{2B_{0}^{\prime }}{B_{0}}\dot{a}%
^{2}+\frac{B_{0}^{\prime }}{B_{0}^{2}}+\frac{C_{0}^{\prime }}{B_{0}C_{0}}%
\right]
\end{equation}%
and%
\begin{equation}
P_{\varphi }=\sqrt{B_{0}}\left[ 2\ddot{a}+\frac{2B_{0}^{\prime }}{B_{0}}\dot{%
a}^{2}+\frac{B_{0}^{\prime }}{B_{0}^{2}}+\frac{D_{0}^{\prime }}{B_{0}D_{0}}%
\right] .
\end{equation}%
The energy conservation identity becomes%
\begin{multline}
\left( S_{;j}^{ij}\overset{i=\tau }{=}\right) \frac{d\sigma }{d\tau }+\left( 
\frac{D^{\prime }}{2D}\left( P_{z}+\sigma \right) +\frac{C^{\prime }}{2C}%
\left( P_{\varphi }+\sigma \right) \right) \frac{da}{d\tau } \\
=-\frac{da}{d\tau }\sigma \left[ \frac{B^{\prime }}{B}-\frac{\zeta }{2}-%
\frac{\zeta ^{\prime }}{\zeta }+\frac{D^{\prime }C^{\prime }}{\zeta DC}%
\right] ,
\end{multline}%
in which 
\begin{equation}
\zeta =\frac{D^{\prime }}{D}+\frac{C^{\prime }}{C}.
\end{equation}%
The potential of the motion of the throat $V\left( a\right) $ reduces to 
\begin{equation}
V\left( a\right) =\frac{1}{B}-\left( \frac{\sigma }{\zeta }\right) ^{2}
\end{equation}%
whose second derivative at point $a=a_{0}$ becomes 
\begin{widetext}
\begin{multline}
V_{0}^{\prime \prime }=\frac{C_{0}^{\prime }\left\{ \left[ \left(
2B_{0}D_{0}^{\prime \prime }-B_{0}^{\prime }D_{0}^{\prime }\right)
D_{0}-2B_{0}D_{0}^{\prime 2}\right] C_{0}^{2}+D_{0}^{2}\left(
2B_{0}C_{0}^{\prime \prime }-B_{0}^{\prime }C_{0}^{\prime }\right)
C_{0}-2D_{0}^{2}C_{0}^{\prime 2}B_{0}\right\} }{2D_{0}B_{0}^{2}\left(
D_{0}^{\prime }C_{0}+C_{0}^{\prime }D_{0}\right) C_{0}^{2}}\phi _{0}^{\prime
} \\
+\frac{D_{0}^{\prime }\left\{ \left[ \left( 2D_{0}D_{0}^{\prime \prime
}-2D_{0}^{\prime 2}\right) C_{0}^{2}+2D_{0}^{2}C_{0}C_{0}^{\prime \prime
}-2D_{0}^{2}C_{0}^{\prime 2}\right] B_{0}-C_{0}D_{0}B_{0}^{\prime }\left(
D_{0}^{\prime }C_{0}+C_{0}^{\prime }D_{0}\right) \right\} }{%
2C_{0}B_{0}^{2}\left( D_{0}^{\prime }C_{0}+C_{0}^{\prime }D_{0}\right)
D_{0}^{2}}\psi _{0}^{\prime } \\
+\frac{2D_{0}\left( B_{0}B_{0}^{\prime \prime }-\frac{3}{2}B_{0}^{\prime
2}\right) D_{0}^{\prime }C_{0}^{2}-2B_{0}^{2}D_{0}C_{0}^{\prime
2}D_{0}^{\prime }}{2D_{0}C_{0}B_{0}^{3}\left( D_{0}^{\prime
}C_{0}+C_{0}^{\prime }D_{0}\right) } \\
+\frac{C_{0}\left[ \left( 2B_{0}B_{0}^{\prime \prime }C_{0}^{\prime
}-3C_{0}^{\prime }B_{0}^{\prime 2}\right) D_{0}^{2}+\left( \left[
2C_{0}^{\prime \prime }D_{0}^{\prime }+2D_{0}^{\prime \prime }C_{0}^{\prime }%
\right] B_{0}^{2}-2C_{0}^{\prime }B_{0}B_{0}^{\prime }D_{0}^{\prime }\right)
D_{0}-2C_{0}^{\prime }B_{0}^{2}D_{0}^{\prime 2}\right] }{%
2D_{0}C_{0}B_{0}^{3}\left( D_{0}^{\prime }C_{0}+C_{0}^{\prime }D_{0}\right) }
\end{multline}%
\end{widetext}in which all functions are calculated at $a=a_{0}$ while $\psi
_{0\text{ }}^{\prime }=\left. \frac{d\psi }{d\sigma }\right\vert _{\sigma
_{0}}$ and $\phi _{0\text{ }}^{\prime }=\left. \frac{d\phi }{d\sigma }%
\right\vert _{\sigma _{0}}.$

\subsection{Stability of the cylindrical TSW with positive cosmological
constant}

In \cite{R}, Richarte introduced a cylindrical wormhole based on the
spacetime in the presence of a cosmic string in vacuum and outside the core
of the string which means $r>r_{core}$ where the bulk metric functions are
given by (for a detailed work see \cite{R})%
\begin{equation}
B\left( a\right) =\cos ^{\frac{4}{3}}\tilde{a}
\end{equation}%
\begin{equation}
C\left( a\right) =\frac{4\delta ^{2}}{3\Lambda }\frac{\sin ^{2}\tilde{a}}{%
\cos ^{\frac{2}{3}}\tilde{a}}
\end{equation}%
and%
\begin{equation}
D\left( a\right) =1.
\end{equation}%
Here $\tilde{a}=\frac{\sqrt{3\Lambda }}{2}a$, $\delta $ is a parameter
related to the deficit angle explicitly given in \cite{BL} and $\Lambda $ is
the cosmological constant.

An explicit calculation of $V_{0}^{\prime \prime }$ yields

\begin{multline}
V_{0}^{\prime \prime }=-\frac{2\Lambda }{3\cos ^{\frac{10}{3}}\tilde{a}%
_{0}\sin ^{2}\tilde{a}_{0}} \\
\left[ \left( \beta _{2}+1\right) \sin ^{4}\tilde{a}_{0}+\frac{3}{2}\left(
1-3\beta _{2}\right) \sin ^{2}\tilde{a}_{0}+\frac{9}{4}\beta _{2}\right] .
\end{multline}

The EoS is a linear gas (LG) in which $\psi ^{\prime }\left( \sigma \right)
=\beta _{1}$ and $\phi ^{\prime }\left( \sigma \right) =\beta _{2}$ where $%
\beta _{1}$ and $\beta _{2}$ are two constant parameters. In order to have a
stable TSW, $V_{0}^{\prime \prime }$ must be positive. With positive
cosmological constant, ultimately, the condition of stability reduces to%
\begin{equation}
\left( \beta _{2}+1\right) \sin ^{4}\tilde{a}_{0}+\frac{3}{2}\left( 1-3\beta
_{2}\right) \sin ^{2}\tilde{a}_{0}+\frac{9}{4}\beta _{2}<0.
\end{equation}%
In Fig. 2 we show the regions of stability in a frame of $\beta _{2}$ versus 
$\tilde{a}.$

\begin{figure}[tbp]
\includegraphics[width=60mm,scale=0.7]{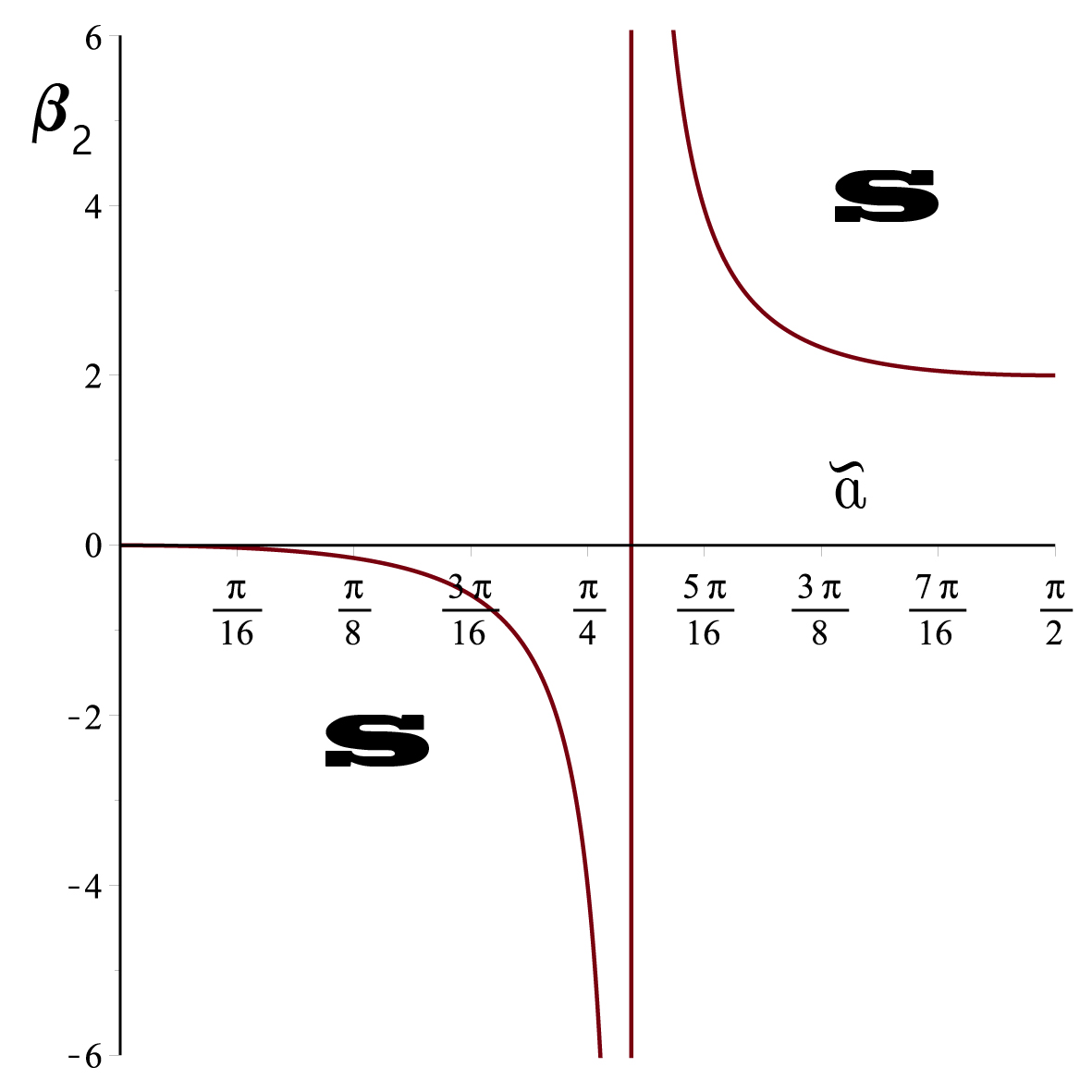}
\caption{Stability of TSW supported by LG in terms of $\tilde{a}_{0}$ and $%
\protect\beta _{2}$. This figure is particularly from Eq. (50).}
\end{figure}

\subsection{Stability of the Brans-Dicke cylindrical TSW}

In \cite{ES2}, Eiroa and Simone presented TSW in Einstein-Brans-Dicke (EBD)
theory. The corresponding metric functions are given by%
\begin{equation}
B\left( a\right) =a^{2d\left( d-n\right) +\left[ \omega \left( n-1\right) +2n%
\right] \left( n-1\right) }
\end{equation}%
\begin{equation}
C\left( a\right) =W_{0}^{2}a^{2\left( n-d\right) }
\end{equation}%
and%
\begin{equation}
D\left( a\right) =a^{2d}.
\end{equation}%
Herein $d$ and $n$ are integration constants such that the scalar field of
the BD theory is given in terms of $n$ as%
\begin{equation}
\phi =\phi _{0}a^{1-n}.
\end{equation}%
Also $\omega >-3/2$ is a free parameter in BD theory while $W_{0}\in 
\mathbb{R}
$. To study the stability of the TSW in this framework, we again consider a
LG for the EoS which means $\psi ^{\prime }=\beta _{1}$ and $\phi ^{\prime
}=\beta _{2}.$ The master equation (45) admits 
\begin{widetext}
\begin{equation}
V_{0}^{\prime \prime }=\frac{2\left( \frac{\Omega }{2}+1+d\left( d-n\right)
\right) \left[ \left( d-\beta _{2}\right) n^{2}+\left( \left[ \beta
_{2}-\beta _{1}-2\right] d-\frac{\Omega }{2}-d^{2}\right) n+2d^{2}\right] }{%
na_{0}^{2d\left( d-n\right) +\Omega +2}}
\end{equation}%
in which $\Omega =$ $\left[ \omega \left( n-1\right) +2n\right] \left(
n-1\right) .$ Imposing $V_{0}^{\prime \prime }>0,$ is equivalant to 
\begin{equation}
\left( \frac{\Omega }{2}+1+d\left( d-n\right) \right) \left[ \left( d-\beta
_{2}\right) n^{2}+\left( \left[ \beta _{2}-\beta _{1}-2\right] d-\frac{%
\Omega }{2}-d^{2}\right) n+2d^{2}\right] >0.
\end{equation}
\end{widetext}This final form of the stability involves too many free
parameters which always renders possible to find some set(s) of parameters
to make the TSW stable. In particular case one can go further to find a more
specific relation.

\subsection{BD solution with a magnetic field}

In \cite{ES2}, in addition to the vacuum metric, the authors considered the
TSWs in cylindrically symmetric BD solution with a magnetic field which was
introduced in \cite{BD}. Based on \cite{ES2,BD} the metric functions are
given by%
\begin{equation}
B\left( a\right) =a^{2d\left( d-n\right) +\Omega }\left(
1+c^{2}a^{-2d+n+1}\right) ^{2},
\end{equation}%
\begin{equation}
C\left( a\right) =\frac{W_{0}^{2}a^{2\left( n-d\right) }}{\left(
1+c^{2}a^{-2d+n+1}\right) ^{2}},
\end{equation}%
and%
\begin{equation}
D\left( a\right) =a^{2d}\left( 1+c^{2}a^{-2d+n+1}\right) ^{2}.
\end{equation}%
As before, $d$ and $n$ are two integration constants and $c$ represents the
magnetic field strength. In the case of BD with the magnetic field the 
\textit{areal flare-out condition} is trivially satisfied but to have 
\textit{radial flare-out condition} satisfied one must consider $c^{2}\left(
d-1\right) a^{-2d+n+1}+n-d>0.$ Clearly when $c=0$ we get the conditions for
the vacuum solution which becomes $n>d.$ Keeping in mind these conditions we
impose $V_{0}^{\prime \prime }>0.$ This in turn yields a very complicated
expression which we refrain to add here but instead we remark that for a
case with $d=n=1$ and $\beta _{1}=\beta _{2}=\beta $ it becomes 
\begin{equation}
V_{0}^{\prime \prime }=-\frac{2\beta }{a^{2}\left( 1+c^{2}\right) ^{2}}
\end{equation}%
which is clearly positive if $\beta <0.$ We note that with our specific
setting only the \textit{areal flare-out condition} is satisfied leaving the
radial flare-out condition open.

\section{Conclusion}

TSWs are considered in cylindrical symmetry where the metric functions rely
entirely on the radial Weyl coordinate. Such spacetimes may not be
asymptotically flat in general so that we expect deviations from the
spherically symmetric counterparts. The source to support the TSW is exotic.
Stability analysis in radial direction is worked out in detail and a master
equation is obtained for an effective potential. This is summarized as $%
V^{\prime \prime }\left( a_{0}\right) >0,$ which turns out to be a tedious
equation for a generic cylindrically symmetric metric. For specific
examples, however, such as Levi-Civita, Brans-Dicke with magnetic fields and
similar cases the stability equation becomes tractable. Parametric plots of
the stability regions can be obtained without much effort. Since our case is
a generic one all known cylindrically symmetric TSW solutions to date can be
cast into our format. Finally we would like to add that in this work we have
only considered the EoS of the fluid which supports the TSW to be LG. Other
possibilities which have been considered so far for the spherical cases such
as, Chaplygin Gas (CG), Generalized Chaplygin Gas (GCG), Modified
Generalized Chaplygin Gas (MGCG) and Logarithmic Gas (LG) are open problems
to be considered \cite{ZHM}.

\bigskip

\end{document}